\documentclass{llncs} %

\usepackage{float}
\usepackage{graphicx}

\usepackage{wrapfig}
\usepackage[mathscr]{euscript}
\usepackage[labelfont=bf]{caption}
\usepackage{color}
\usepackage{amsmath} 
\usepackage{mathtools} 
\usepackage{amsfonts} 
\usepackage{amssymb}

\usepackage{array}
\usepackage{booktabs}
\usepackage{rotating}
\usepackage{multirow}
\usepackage{multicol}
\usepackage{lscape}
\usepackage{subfig}

\usepackage[export]{adjustbox}
\usepackage{color}

\usepackage[colorlinks=true,linkcolor=blue,citecolor=blue]{hyperref}

\definecolor{brightpink}{rgb}{1.0, 0.0, 0.5}
\definecolor{crimson}{rgb}{0.86, 0.08, 0.24}
\definecolor{flame}{rgb}{0.89, 0.35, 0.13}

\usepackage{algorithm,algorithmicx,algpseudocode}

\usepackage{color}

\usepackage{tikz,xcolor,hyperref}

\definecolor{lime}{HTML}{A6CE39}
\DeclareRobustCommand{\orcidicon}{
	\begin{tikzpicture}
	\draw[lime, fill=lime] (0,0) 
	circle [radius=0.16] 
	node[white] {{\fontfamily{qag}\selectfont \tiny ID}};
	\draw[white, fill=white] (-0.0625,0.095) 
	circle [radius=0.007];
	\end{tikzpicture}
	\hspace{-2mm}
}

\foreach \x in {A, ..., Z}{\expandafter\xdef\csname orcid\x\endcsname{\noexpand\href{https://orcid.org/\csname orcidauthor\x\endcsname}
			{\noexpand\orcidicon}}
}
			
\definecolor{darkgreen}{rgb}{0.53, 0.66, 0.42}

\begin{document}

\title{GSR-Net: Graph Super-Resolution Network for Predicting High-Resolution from Low-Resolution Functional Brain Connectomes}

\titlerunning{Short Title}  

\author{Megi Isallari\index{Isallari, Megi} \and Islem Rekik\index{Rekik, Islem}\orcidA{}\thanks{{Corresponding author: irekik@itu.edu.tr, \url{http://basira-lab.com}.} This work is accepted for publication in the Machine Learning in Medical Imaging workshop Springer proceedings, in conjunction with MICCAI 2020.}}

\authorrunning{M. Isallari et al.}  

\institute{BASIRA Lab, Faculty of Computer and Informatics, Istanbul Technical University, Istanbul, Turkey \vspace{-6ex}}
\maketitle              

\begin{abstract}

Catchy but rigorous deep learning architectures were tailored for image super-resolution (SR), however, these fail to generalize to non-Euclidean data such as brain connectomes.  Specifically, building generative models for \emph{super-resolving a low-resolution brain connectome} at a higher resolution (i.e., adding new graph nodes/edges) remains unexplored —although this would circumvent the need for costly data collection and manual labelling of anatomical brain regions (i.e. parcellation). To fill this gap, we introduce GSR-Net (Graph Super-Resolution Network), the first super-resolution framework operating on graph-structured data that generates high-resolution brain graphs from low-resolution graphs. \emph{First}, we adopt a U-Net like architecture based on graph convolution, pooling and unpooling operations specific to non-Euclidean data. However, unlike conventional U-Nets where graph nodes represent samples and node features are mapped to a low-dimensional space (encoding and decoding node attributes or sample features), our GSR-Net operates \emph{directly} on a single connectome: a fully connected graph where conventionally, a node denotes a brain region, nodes have no features, and edge weights denote brain connectivity strength between two regions of interest (ROIs).  In the absence of original node features, we initially assign identity feature vectors to each brain ROI (node) and then leverage the learned local receptive fields to learn node feature representations. Specifically, for each ROI, we learn a node feature embedding by locally averaging the features of its neighboring nodes based on their connectivity weights. \emph{Second}, inspired by spectral theory, we break the symmetry of the U-Net architecture by topping it up with a graph super-resolution (GSR) layer and two graph convolutional network layers to predict a HR (high-resolution) graph while preserving the characteristics of the LR (low-resolution) input. Our proposed GSR-Net framework outperformed its variants for predicting high-resolution brain functional connectomes from low-resolution connectomes. Our Python GSR-Net code is available on BASIRA GitHub at \url{https://github.com/basiralab/GSR-Net}.

\end{abstract}

\section{Introduction}

Remarkable progress in diagnosing brain disorders and exploring brain anatomy has been made using neuroimaging modalities (such as MRI (magnetic resonance imaging) or DTI (diffusion tensor imaging)). Recent advances in ultra-high field (7 Tesla) MRI help show fine-grained variations in brain structure and function. However, MRI data at submillimeter resolutions is very scarce due to the limited number and high cost of the ultra-high field scanners. To circumvent this issue, several works explored the prospect of super-resolution to map a brain intensity image of low resolution to an image of higher resolution \cite{bahrami20177t,chen2018brain,ebner2020automated}. In recent years, advances in deep learning have inspired a multitude of works in image super-resolution ranging from the early approaches using Convolutional Neural Networks (CNN) (e.g. SRCNN \cite{dong2014image}) to the state-of-the-art methods such as Generative Adversarial Nets (GAN) (e.g. SRGAN \cite{ledig2016photorealistic}). For instance,  \cite{inproceedings} used Convolutional Neural Networks to generate 7T-like MRI images from 3T MRI and more recently, \cite{Lyu_2020} used ensemble learning to synergize high-resolution GANs of MRI differentially enlarged with complementary priors. While a significant number of image super-resolution methods have been proposed for MRI super-resolution, super-resolving brain connectomes (i.e., brain graphs) remains largely unexplored. Typically, a brain connectome is the product of a very complex neuroimage processing pipeline that integrates MRI images into pre-processing and analysis steps from skull stripping to cortical thickness, tissue segmentation and registration to a brain atlas \cite{Bassett:2017}. To generate brain connectomes at different resolutions, one conventionally uses image brain atlas (template) to define the parcellation of the brain into $N$ (depending on the resolution) anatomical regions of interest (ROIs). A typical brain connectome is comprised of $N$ nodes where a node denotes a brain ROI and edge weights denote brain connectivity strength between two ROIs (e.g., correlation between neural activity or similarity in brain morphology) \cite{Fornito:2015,Heuvel:2019}. However, this process has two main drawbacks: (1)  the computational time per subject is very high and (2) pre-processing steps such as registration and label propagation are highly prone to variability and bias \cite{qi2015influence,bressler2010large}. 

\emph{Alternatively, given a low-resolution (LR) connectome, one can devise a systematic method to automatically generate a high-resolution (HR) connectome  and thus circumvent the need for costly neuroimage processing pipelines.} However, such a method would have to address two major challenges. \emph{First}, standard downsampling/upsampling techniques are not easily generalizable to non-Euclidean data due to the complexity of network data. The high computational complexity, low parallelizability, and inapplicability of machine learning methods to geometric data render image super-resolution algorithms ineffective \cite{cui}. \emph{Second}, upsampling (super-resolution) in particular is a notoriously ill-posed problem since the LR connectome can be mapped to a variety of possible solutions in HR space. Furthermore, while unpooling (deconvolution) is a recurring concept in graph embedding approaches, it typically focuses on graph embedding reconstruction rather than in the expansion of the topology of the graph \cite{unets}. Two recent pioneering works have tackled the problem of graph super-resolution \cite{Cengiz:2019,Mhiri:2020}, however both share the dichotomized aspect of the engineered learning-based GSR framework, which is composed of independent blocks that cannot co-learn together to better solve the target super-resolution problem. Besides, both resort to first vectorizing LR brain graphs in  the beginning of the learning process, thereby spoiling the rich topology of the brain as a connectome.

To address these limitations, we propose GSR-Net: the first geometric deep learning framework that attempts to solve the problem of predicting a high-resolution connectome from a low-resolution connectome. The key idea of GSR-Net can be summarized in three fundamental steps: (i) learning feature embeddings for each brain ROI (node) in the LR connectome, (ii) the design of a graph super-resolution operation that predicts an HR connectome from the  LR connectivity matrix and feature embeddings of the LR connectome computed in (i), (iii) learning node feature embeddings for each node in the super-resolved (HR) graph obtained in (ii).    
First, we adopt a U-Net like architecture and introduce the Graph U-Autoencoder. Specifically, we leverage the Graph U-Net proposed in \cite{unets}: an encoder-decoder architecture based on graph convolution, pooling and unpooling operations that specifically work on non-Euclidean data. However, as most graph embedding methods, the Graph U-Net focuses on typical graph analytic tasks such as link prediction or node classification rather than super-resolution. Particularly, the conventional Graph U-Net is a \emph{node-focused} architecture where a node $\mathit{n}$ represents a sample and mapping the node $\mathit{n}$ to an $\mathit{m}$-dimensional space (i.e., simpler representation) depends on the node and its attributes \cite{GNNmodel}.

\begin{figure}[ht!]
\centering
\includegraphics[width=12.5cm]{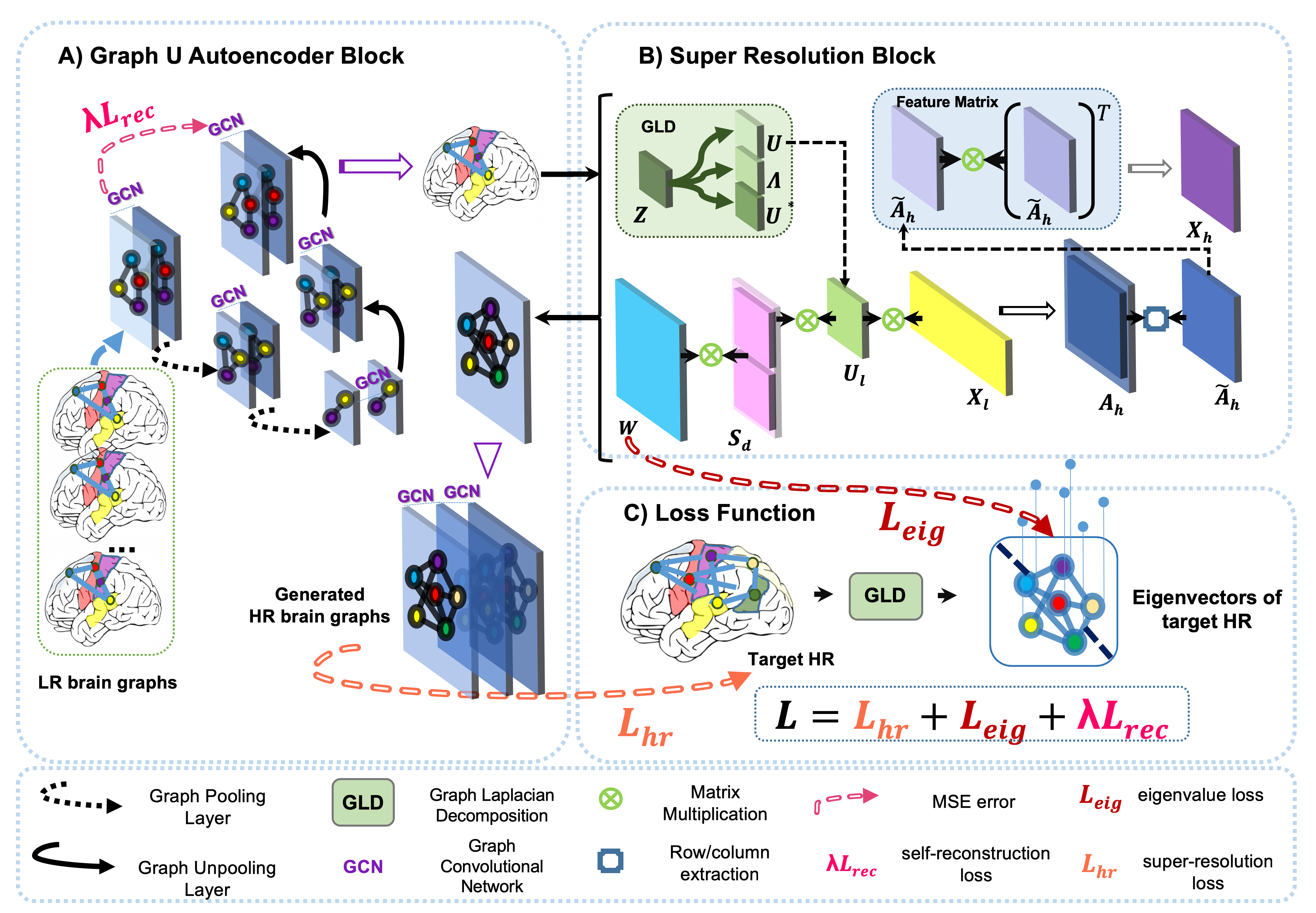}
\caption{\emph{Proposed framework of Graph Super-Resolution Network (GSR-Net) for super-resolving low-resolution brain connectomes.} \textbf{(A) Graph U-Autoencoder Block.} Our Graph U-Autoencoder is built by stacking two encoding modules and two decoding modules. An encoding module
contains a graph pooling layer and a graph convolutional network (GCN) and its inverse operation is a decoding module comprised of a graph unpooling layer and a GCN. Here, we integrate a \emph{self-reconstruction loss} $\textcolor{brightpink}{\mathcal{L}_{rec}}$ that guides the learning of node feature embeddings for each brain ROI in the LR connectome. \textbf{(B) Super Resolution Block.} The GSR Layer super-resolves both the topological structure of the LR connectome (connectivity matrix $\mathbf{A}_{l}$) and the feature matrix of the LR connectome ($\mathbf{X}_{l}$). To super-resolve $\mathbf{A}_{l}$, we propose the layer-wise propagation rule  $\tilde{{\mathbf{A}}}_{h} =  \mathbf{W} \mathbf{S}_{d}{\mathbf{U}_{0}}^{*}\mathbf{Z}_{l}$, where $\mathbf{W}$ is a matrix of trainable filters that we enforce to match the eigenvector matrix of the HR graph via an \emph{eigen-decomposition loss} $\textcolor{crimson}{\mathcal{L}_{eig}}$, $\mathbf{S_{d}}$ is the concatenation of two identity matrices, $\mathbf{U}_{0}$ is the eigenvector matrix of $\mathbf{A}_{l}$ and $\mathbf{Z}_{l}$ is the matrix of node feature embeddings of the LR brain graph generated in \textbf{(A)}. The propagation rule for the feature matrix super-resolution is: $\tilde{\mathbf{X}}_{h} = \tilde{{\mathbf{A}}}_{h}{\tilde{{\mathbf{A}}}_{h}}^T$. \textbf{(C) Loss function.} Our GSR-Net loss comprises a self-reconstruction loss $\textcolor{brightpink}{\mathcal{L}_{rec}}$, super-resolution loss $\textcolor{flame}{\mathcal{L}_{hr}}$ and eigen-decomposition loss $\textcolor{crimson}{\mathcal{L}_{eig}}$ to optimize learning the predicted HR connectome from a LR connectome.} 
\label{GSRNet}
\end{figure}

Our Graph U-Autoencoder on the other hand, is a \emph{graph-focused} architecture where a sample is represented by a \emph{connectome}: a fully connected graph where conventionally, nodes have no features and edge weights denote brain connectivity strength between two nodes. We unify both these concepts by learning a mapping of the node $\mathit{n}$ to an $\mathit{m}$-dimensional space that translates the topological relationships between the nodes in the connectome as node features. Namely, we initially assign identity feature vectors to each brain ROI and we learn node feature embeddings by locally averaging the features of its neighboring nodes based on their connectivity weights.  Second, we break the symmetry of the U-Net architecture by adding a GSR layer to generate an HR connectome from the node feature embeddings of the LR connectome learned in the Graph U-Autoencoder block. Specifically, in our GSR block, we propose a layer-wise propagation rule for super-resolving low-resolution brain graphs, rooted in spectral graph theory. Third,  we stack two additional graph convolutional network layers to learn node feature embeddings for each brain ROI in the super-resolved graph.

\section{Proposed GSR-Net for Brain Connectome Super-Resolution}

\textbf{Problem Definition. } A connectome can be represented as \( \mathbf{C}=\{ \mathbf{V},\mathbf{E},\mathbf{X} \}\) where \(\mathbf{V}\) is a set of nodes and \(\mathbf{E}\) is a set of edges connecting pairs of nodes. The network nodes are defined as brain ROIs. The connectivity (adjacency) matrix \(\mathbf{A}\) is an \(N \times N\) matrix (\(N\) is the number of nodes), where $\mathbf{A}_{ij}$  denotes the connectivity weight between two ROIs $i$ and $j$ using a specific metric (e.g., correlation between neural activity or similarity in brain morphology). 
Let \(\mathbf{X} \in \mathbb{R}^{N \times F}\) denote the feature matrix where \(N\) is the number of nodes and $F$  is the number of features (i.e., connectivity weights) per node.  Each training subject \(s\) in our dataset is represented by two connectivity matrices in LR and HR domains denoted as \(\mathbf{C}_{l} =  \{ \mathbf{V}_{l}, \mathbf{E}_{l}, \mathbf{X}_{l} \} \) and \(\mathbf{C}_{h} = \{ \mathbf{V}_{h}, \mathbf{E}_{h}, \mathbf{X}_{h} \}\), respectively. Given a brain graph \(\mathbf{C}_{l}\), our objective is to learn a mapping $f: (\mathbf{A}_{l},\mathbf{X}_{l}) \mapsto (\mathbf{A}_{h},\mathbf{X}_{h})$, which maps $\mathbf{C}_{l}$ onto $\mathbf{C}_{h}$.

\textbf{Overall Framework.} In \textbf{Fig}~\ref{GSRNet}, we illustrate the proposed GSR-Net architecture including:  \textbf{(i)} an asymmetric graph U-Autoencoder to learn the feature embeddings matrix \(\mathbf{Z}_{l}\) for a LR brain graph by $f_{l}: (\mathbf{A}_{l}, \mathbf{X}_{l}) \mapsto \mathbf{Z}_{l}$,  \textbf{(ii)} a graph super-resolution (GSR) layer mapping LR graph embeddings \(\mathbf{Z}_{l}\) and the LR connectivity matrix to a HR feature matrix and connectivity matrix  by $f_{h}: (\mathbf{A}_{l}, \mathbf{Z}_{l}) \mapsto (\mathbf{\tilde{A}}_{h}, \mathbf{\tilde{X}}_{h})$,  \textbf{(iii)} learning the HR feature embeddings \(\mathbf{Z}_{h}\) by stacking two graph convolutional layers as $f_{z}: (\mathbf{\tilde{A}}_{h}, \mathbf{\tilde{X}}_{h}) \mapsto \mathbf{Z}_{h}$, and \textbf{(iv)} computing the loss function $\mathbf{\mathscr{L}}$.

\textbf{1. Graph U-Autoencoder.} U-Net architectures have long achieved state-of-the-art performance in various tasks thanks to their encoding-decoding nature for high-level feature extraction and embedding.  In the first step of our GSR-Net, we adopt the concept of Graph U-Nets \cite{unets} based on learning node representations from node attributes and we extend this idea to learning node representations from topological relationships between nodes. To learn node feature embeddings of a given LR connectome \( \mathbf{C}_{l}=\{ \mathbf{V}_{l},\mathbf{E}_{l},\mathbf{X}_{l} \}\),  we propose a Graph U-Autoencoder comprising of a Graph U-Encoder and a Graph U-Decoder. 

\textbf{\emph{Graph U-Encoder}}. The Graph U-Encoder inputs the adjacency matrix $\mathbf{A}_l \in \mathbb{R}^{N \times N}$ of  \( \mathbf{C}_{l}=\{ \mathbf{V}_{l},\mathbf{E}_{l},\mathbf{X}_{l} \}\) (N is the number of nodes of $\mathbf{C}_l$) as well as the feature matrix capturing the node content of the graph \(\mathbf{X}_{l} \in \mathbb{R}^{N \times F}\).  In the absence of original node features, we assign an identity matrix \(\mathbf{I}_{N} \in \mathbb{R}^{N \times N}\) to the feature matrix \(\mathbf{X}_{l}\) , where the encoder is only informed of the identity of each node.
We build the Graph U-Encoder by stacking multiple encoding modules, each containing a graph pooling layer followed by a graph convolutional layer. Each encoding block is intuitively expected to encode high-level features by downsampling the connectome and aggregating content from each node's local topological neighborhood. However, as a \emph{graph-focused} approach where the sample is represented by a connectome and the connectome's nodes are featureless, our Graph U-Encoder defines the notion of locality by edge weights rather than node features. Specifically, the pooling layer adaptively selects a few nodes to form a smaller brain graph in order to increase the local receptive field and for each node, the GCN layer aggregates (locally averages) the features of its neighboring nodes based on their connectivity weights. 

\emph{Graph Pooling Layer}. The layer's propagation rule can be defined as follows:

$v = \mathbf{X}_l^{(l)} u^{(l)} /\parallel u^{(l)}\parallel; indices = rank(v,k);        \tilde{v} = sigmoid(v(indices));\\
\tilde{\textbf{X}_l}^{(l)} = \textbf{X}_l^{(l)}(indices,:); \textbf{A}_l^{(l+1)}=\textbf{A}_l^{(l)}(indices,indices); \textbf{X}_l^{(l+1)}=\tilde{\textbf{X}_l}^{(l)}\odot(\tilde{v} 1^{T}_F) $

The graph pooling layer adaptively selects a subset of nodes to form a new smaller graph
based on their scalar projection values on a trainable projection vector $u$.  First, we find the scalar projection of \(\mathbf{X}_{l}\) on \(u\) which computes a one-dimensional \(v\) vector, where \(v_{i}\) is the scalar projection of each node on vector \(u\). We find the k-largest values in \(v\) which are then saved as the indices of the nodes selected for the new downsampled graph. According to the indices found, we extract the feature matrix rows for each node selected ($\mathbf{X}_l^{(l)}(indices,:)$) as well as the respective adjacency matrix rows and columns to obtain the adjacency matrix of the downsampled graph: $\mathbf{A}_l^{(l+1)} = \mathbf{A}_l^{(l)}(indices,indices)$. Hence, this reduces the graph size from $N$ to $k$ : $\mathbf{A}_l^{(l+1)} \in \mathbb{R}^{k \times k}$. In the end, by applying a sigmoid mapping to the projection vector \(v\), we obtain the gate vector \(\tilde{v} \in \mathbb{R}^k\) which we multiply with \(\mathbf{1}^{T}_F \) (one-dimensional vector with all \(F\) elements equal to 1). The product $\tilde{v} 1^{T}_F$ is then multiplied element-wise with \(\tilde{\mathbf{X}}_l^{(l)} \) to control information of the selected nodes and obtain the new feature matrix of the downsampled graph $\mathbf{X}_l^{(l+1)} \in \mathbb{R}^{k \times F}$.

\emph{\textbf{Graph U-Decoder}}. Similarly to Graph U-Encoder, Graph U-Decoder is built by stacking multiple decoding modules, each comprising a graph unpooling layer followed by a graph convolutional layer. Each decoding module acts as the inverse operation of its encoding counterpart by gradually upsampling and aggregating neighborhood information for each node.\\ \emph{Graph Unpooling Layer.} The graph unpooling layer retracts the graph pooling operation by relocating the nodes in their original positions according to the saved indices of the selected nodes in the pooled graph. Formally, we write $\mathbf{X}^{(l+1)} = relocate(\mathbf{0}_{N \times F},\mathbf{X}_{l}^{(l)},indices)$, where \(\mathbf{0}_{N \times F}\) is the reconstructed feature matrix of the new graph (initially the feature matrix is empty) . \(\mathbf{X}_{l}^{(l)} \in \mathbb{R}^{k \times F}\) is the feature matrix of the current downsampled graph and the $relocate$ operation assigns row vectors in $\mathbf{X}_{l}^{(l)}$ into $\mathbf{0}_{N \times F}$ feature matrix according to their corresponding indices stored in $indices$.  

\textbf{\textit{Graph U-Autoencoder for super-resolution.}} Next, we introduce our Graph U-Autoencoder which first includes a GCN to learn an initial node representation of the LR connectome. This first GCN layer 
takes as input \((\mathbf{A}_{l},\mathbf{X}_{l})\) and outputs  \(\mathbf{Z}_{0} \in \mathbb{R}^{N \times NK}\) : a node feature embedding matrix with $NK$ number of features per node where \(K\) is the factor by which the resolution increases when we predict the HR graph from a LR graph (\(F\) is specifically chosen to be \(NK\) for reasons we explore in greater detail in the next section). The transformation can be defined as follows: $\mathbf{Z}_{0} = \sigma(\hat{\mathbf{D}}^{-\frac{1}{2}}\hat{\mathbf{A}}\hat{\mathbf{D}}^{-\frac{1}{2}}\mathbf{X}_l\mathbf{W}_l)$, where $\hat{\mathbf{D}}$ is the diagonal node degree matrix, $\hat{\mathbf{A}} = \mathbf{A} + \mathbf{I}$ is the adjacency matrix with added self-loops and $\sigma$ is the activation function. $\mathbf{W}_l$ is a matrix of trainable filter parameters to learn. Next, we apply two encoding blocks followed by two decoding blocks outputting $\mathbf{Z}_{l} \in \mathbb{R}^{N \times NK}$:  $\mathbf{Z}_l = GraphUAutoencoder(\hat{\mathbf{A}_l}, \mathbf{Z}_0).$

\textbf{\emph{Optimization}}. To improve and regularize the training of our graph autoencoder model such that the LR connectome embeddings preserve the topological structure $\mathbf{A}_{l}$ and node content information $\mathbf{X}_{l}$ of the original LR connectome, we enforce the learned LR node feature embedding $\mathbf{Z}_l$ to match the initial node feature embedding of the LR connectome $\mathbf{Z}_{0}$. In our loss function we integrate a \emph{self-reconstruction regularization term} which minimizes the mean squared error (MSE)  between the node representation \(\mathbf{Z}_{0}\) and the output of the Graph U-Autoencoder \(\mathbf{Z}_{l}\): $\textcolor{brightpink}{\mathcal{L}_{rec} = \lambda\frac{1}{N}\sum_{i=1}^{N} ||{\mathbf{Z}_{0}}_{i} - {\mathbf{Z}_{l}}_{i}||_2^2}$.

\textbf{2. Proposed GSR layer.} Super-resolution plays an important role in grid-like data but standard image operations are not directly applicable to graph data. In particular, there is no spatial locality information among nodes in graphs. In this section, we present a mathematical formalization of the GSR Layer, which is the key operation for predicting a high-resolution graph \(\mathbf{C}_{h}\) from the low-resolution brain graph \(\mathbf{C}_{l}\).  Recently, \cite{Tanaka} proposed a novel upsampling method rooted in graph Laplacian decomposition that aims to upsample a graph signal while retaining the frequency domain characteristics of the original signal defined in the time/spatial domain. To define our GSR layer, we leverage the spectral upsampling concept to expand the size of graph while perserving the local information of the node \emph{and} the global structure of the graph using the spectrum of its graph Laplacian.  

Suppose \(\mathbf{L}_{0} \in \mathbb{R}^{N \times N}\) and \(\mathbf{L}_{1} \in \mathbb{R}^{NK \times NK}\) are the graph Laplacians of the original low-resolution graph and high-resolution (upsampled) graph respectively ($K$ is the factor by which the resolution of the graph increases). Given \(\mathbf{L}_{0}\) and \(\mathbf{L}_{1}\),  their respective eigendecompositions are: $\mathbf{L}_{0} = \mathbf{U}_{0} \Lambda {\mathbf{U}_{0}}^{*}, \mathbf{L}_{1} = \mathbf{U}_{1} \Lambda {\mathbf{U}_{1}}^{*}  $ , where \(\mathbf{U}_{0} \in \mathbb{R}^{N \times N}\) and \(\mathbf{U}_{1} \in \mathbb{R}^{NK \times NK}\). In matrix form, our graph upsampling definition can be easily defined as: $x_{u} = \mathbf{U}_{1}\mathbf{S}_{d}{\mathbf{U}_{0}}^{*}x$, where \(\mathbf{S}_{d} = [ \mathbf{I}_{N \times N}   \mathbf{I}_{N \times N}]^T\), $x$ is a signal on the input graph and $x_u$ denotes the upsampled signal. We can generalize the matrix form to a signal \(\mathbf{X}_{l} \in \mathbb{R}^{N \times F}\) with \(F\) input channels (i.e., a \(F\)-dimensional vector for every node) as follows: $\tilde{\mathbf{A}}_{h} =  \mathbf{U}_{1}\mathbf{S}_{d}{\mathbf{U}_{0}}^{*}\mathbf{X}_{l}$. To generate an \(NK \times NK\) resolution graph, the number of input channels \(F\) of \(\mathbf{X}_{l}\)  should be set to \(NK\). This is why the output of the Graph U-AutoEncoder \(\mathbf{Z}_{l}\) (which is going to be the input \(\mathbf{X}_{l}\) of the GSR Layer) is specified to be of the dimensions: \(N \times NK\).

\emph{\textbf{Super-resolving the graph structure}}. To predict \(\tilde{\mathbf{A}}_{h}\),  we first predict the eigenvectors \(\mathbf{U}_{1}\) of the ground truth high-dimensional \(\mathbf{A}_{h}\). We formalize the learnable parameters in this GSR layer as a matrix \(\mathbf{W} \in \mathbb{R}^{NK \times NK}\) to learn  such that the distance error between the weights and the eigenvectors \(\mathbf{U}_{1}\) of the ground truth high-resolution \(\mathbf{A}_{h}\) is minimized.  Hence, the propagation rule for our layer is: $\tilde{{\mathbf{A}}}_{h} =  \mathbf{W} \mathbf{S}_{d}{\mathbf{U}_{0}}^{*}\mathbf{Z}_{l}$.

\emph{\textbf{Super-resolving the graph node features}}. To super-resolve the feature matrix or assign feature vectors to the new nodes (at this point, the new nodes do not have meaningful representations), we again leverage the concept of translating topological relationships between nodes to node features. By adding new nodes and edges while attempting to retain the characteristics of the original low-resolution brain graph, it is highly probable that some new nodes and edges will remain isolated, which might cause loss of information in the subsequent layers. To avoid this, we initialize the target feature matrix \(\tilde{\mathbf{C}}_{h}\) as follows: $\tilde{\mathbf{X}}_{h}^{(l)} = \tilde{{\mathbf{A}}}_{h}^{(l)}   (\mathbf{\tilde{A}}_{h}^{(l)})^T$. This operation links nodes at a maximum two-hop distance and  increases connectivity between nodes \cite{chepuri2016subsampling}. Each node is then assigned a feature vector that satisfies this property. Notably, both the adjacency and feature matrix are converted to symmetric matrices mimicking realistic predictions:  $\mathbf{\tilde{A}}_{h} = (\mathbf{\tilde{A}}_{h} + {\mathbf{\tilde{A}}_{h}}^{T})/2$ and $\mathbf{\tilde{X}}_{h} = (\mathbf{\tilde{X}}_{h} + {\mathbf{\tilde{X}}_{h}}^{T})/2$.

\textbf{\emph{Optimization.}} To learn trainable filters which enforce the super-resolved connectome's eigen-decomposition to match that of the ground truth HR connectome (i.e., preserving both local and global topologies), we further add the \emph{eigen-decomposition loss}: the MSE between the weights and the eigenvectors \(\mathbf{U}_{1}\) of the ground truth high-resolution $\mathbf{A}_{h}$: 
$\textcolor{crimson}{\mathcal{L}_{eig} = \frac{1}{N} \sum_{i=1}^{N} ||\mathbf{W}_{i} - {\mathbf{U}_{1}}_{i}||_2^2}$ .

\textbf{3. Additional graph embedding layers. } Following the GSR layer, we learn more representative ROI-specific feature embeddings of the super-resolved graph by stacking two additional GCNs: $\mathbf{Z}_h^0 = GCN(\tilde{\mathbf{A}}_h,\tilde{\mathbf{X}}_h)$ and $\mathbf{Z}_h = GCN(\tilde{\mathbf{A}}_h,{\tilde{\mathbf{Z}}_h}^0)$. For each node, these embedding layers aggregate the feature vectors of its neighboring nodes, thus fully translating the connectivity weights to node features of the new super-resolved graph. The output of this third step constitutes the final prediction of the GSR-Net of the HR connectome from the input LR connectome. However, our predictions of the HR graph $\mathbf{Z}_h$ are of size $NK \times NK$  and our target HR graph size might not satisfy such multiplicity rule. In such case, we can add isotropic padding of HR adjacency matrix during the training stage and remove the extra-padding in the loss evaluation step and in the final prediction.

\textbf{\emph{Optimization}}. Our training process is primarily guided by the \emph{super-resolution loss} which minimizes the MSE between our super-resolved brain connectomes and the ground truth HR ones. The total GSR-Net loss function comprises the \emph{self-reconstruction loss}, the \emph{eigen-decomposition loss}, and the \emph{super-resolution loss} and it is computed as follows:

\begin{gather*}
\mathcal{L} =  \textcolor{flame}{\mathcal{L}_{hr}} + \textcolor{crimson}{\mathcal{L}_{eig}} + \lambda \textcolor{brightpink}{\mathcal{L}_{rec}}  \\ =
\textcolor{flame}{\frac{1}{N}\sum_{i=1}^{N} ||{\mathbf{Z}_{h}}_{i} - {\mathbf{A}_{h}}_{i}||_2^2 } + \textcolor{crimson}{\frac{1}{N}\sum_{i=1}^{N} ||\mathbf{W}_{i} - {\mathbf{U}_{1}}_{i}||_2^2} + \lambda \textcolor{brightpink}{\frac{1}{N}\sum_{i=1}^{N} ||{\mathbf{Z}_{0}}_{i} - {\mathbf{Z}_{l}}_{i}||_2^2}
\end{gather*}

\section{Results and Discussion}

\textbf{Connectomic dataset and parameter setting.} We used 5-fold cross-validation to evaluate our framework on 277 subjects  from the Southwest University Longitudinal Imaging Multimodal (SLIM) study \cite{slimstudy}. For each subject, two separate functional brain networks with $160 \times 160$ (LR) and $268 \times 268$ (HR) resolutions were produced using two groupwise whole-brain parcellation approaches proposed in \cite{Dosenbach1358} 
and \cite{SHEN2013403}, respectively.  
Our GSR-Net uses Adam Optimizer with a learning rate of $0.0001$ and the number of neurons in both Graph U-Autoencoder and GCN layers is set to $NK$. We empirically set the parameter $\lambda$ of the self-reconstruction regularization loss to $16$. 

\begin{figure}[htp!]
\centering
{\includegraphics[width=12cm]{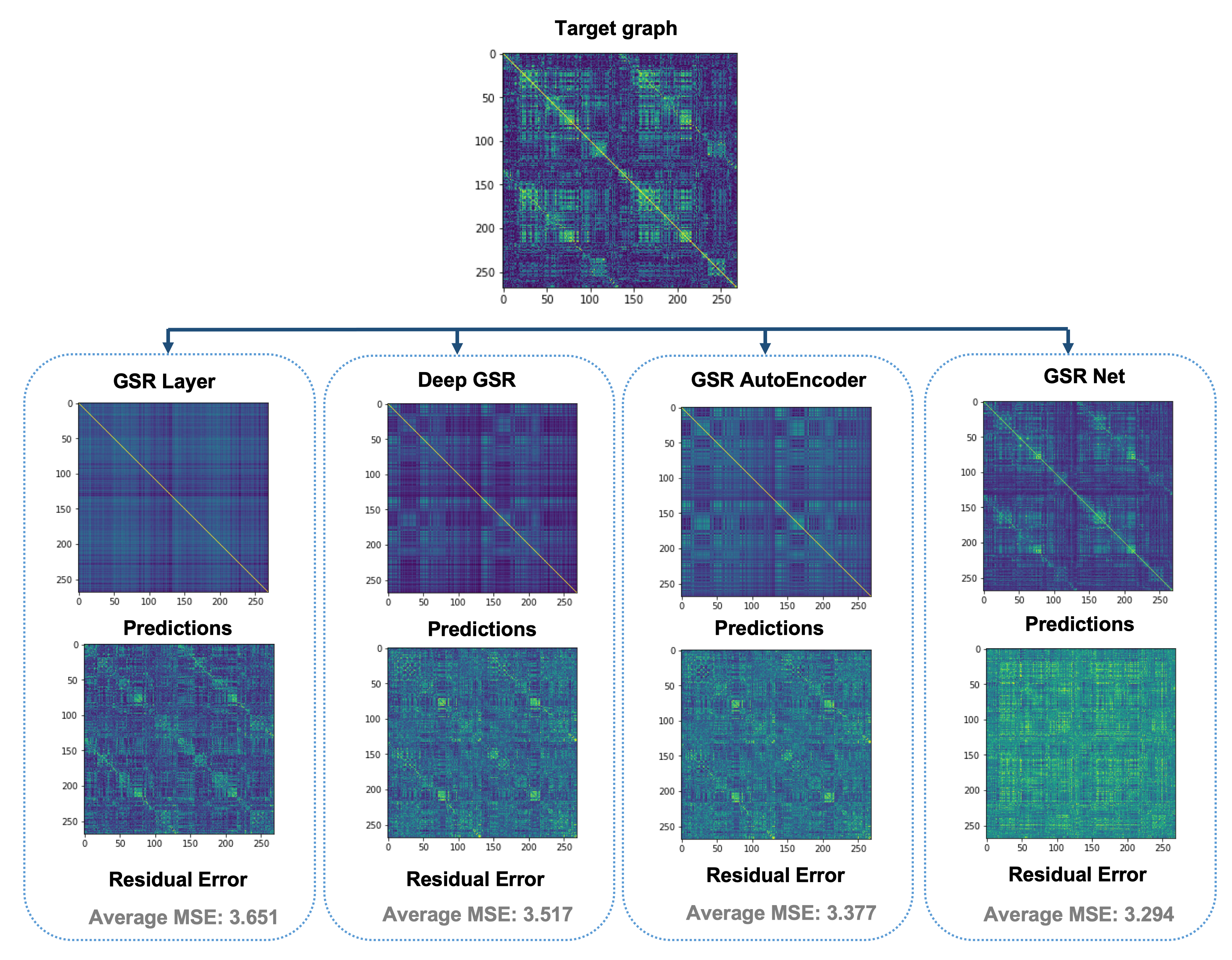}}
\caption{\emph{Comparison between the ground truth HR graph and the predicted HR graph of a representative subject.} We display in \textbf{(A)} the residual error matrix computed using mean squared error (MSE) between the ground truth and predicted super-resolved brain graph. We plot in \textbf{(B)} MSE results for each of the three baseline methods and our proposed GSR-Net.}
\label{fig:2}
\end{figure}

\textbf{Evaluation and comparison methods.} 
We benchmark the performance of our GSR-Net against different baseline methods:
\textbf{(1) GSR Layer:} a variant of GSR-Net where we remove both the graph Autoencoder (\textbf{Fig}~\ref{GSRNet}--A) and the additional graph embedding layers.
\textbf{(2) Deep GSR:} In this variant, first, the node feature embeddings matrix $\mathbf{Z}_l$ of the LR connectome is learned through two GCN layers. Second, this $Z_l$ is inputted to the GSR Layer, and third we learn the node feature embeddings  of the output of the GSR Layer (i.e., the super-resolved graph) leveraging two more GCN layers and a final inner product decoder layer.  
\textbf{(3) GSR-Autoencoder}: a variant of GSR-Net where we remove only the additional GCN layers.
\textbf{Fig}~\ref{fig:2}--B displays the average MSE between the ground truth and predicted HR brain graphs by all methods. Our GSR-Net achieved the best super-resolution performance. 
For a representative subject, we also display the ground truth and predicted HR graphs by all methods along with their residual error. GSR-Net clearly achieves the lowest residual error. Building on this first work, we will further extend our GSR-Net architecture to predict brain connectomes at different resolutions from a low-resolution brain connectome, which can be leveraged in comparative connectomics \cite{van2016comparative} as well as charting the \emph{multi-scale} landscape of brain dysconnectivity in a wide spectrum of disorders \cite{Heuvel:2019}.

\section{Conclusion}

In this paper, we proposed GSR-Net, the first geometric deep learning framework for super-resolving  low-resolution functional brain connectomes. Our method achieved the best graph super-resolution results in comparison with its ablated version and other variants. However, there are a few limitations we need to address. To circumvent the high computational cost of a graph Laplacian,  we can well-approximate the eigenvalue vector by a truncated expression in terms of Chebyshev polynomials \cite{hammond2009wavelets}. Future work includes refining our spectral upsampling theory towards fast computation, enhancing the scalability and interpretability of our GSR-Net architecture with recent advancements in geometric deep learning, and extending its applicability to large-scale multi-resolution brain connectomes \cite{bressler2010large}. Besides, we aim to condition the learning of the HR brain graph by a population-driven connectional brain template \cite{Dhifallah:2020} to enforce the super-resolution of more biologically sound  brain connectomes.

\section{Supplementary material}

We provide three supplementary items on GSR-Net for reproducible and open science:

\begin{enumerate}
	\item A 12-mn YouTube video explaining how GSR-Net works on BASIRA YouTube channel at \url{https://youtu.be/xwHKRxgMaEM}.
	\item GSR-Net code in Python on GitHub at \url{https://github.com/basiralab/GSR-Net}. 
	\item A GitHub video code demo on BASIRA YouTube channel at \url{https://youtu.be/GahVu9NeOIg}. 
\end{enumerate}

\section{Acknowledgement}

This project has been funded by the 2232 International Fellowship for
Outstanding Researchers Program of TUBITAK (Project No:118C288, \url{http://basira-lab.com/reprime/}) supporting I. Rekik. However, all scientific contributions made in this project are owned and approved solely by the authors.

\bibliography{references}
\bibliographystyle{splncs}
\end{document}